\documentclass[12pt]{article}
\usepackage{amsmath}
\usepackage{amssymb}
\usepackage{epsfig}

\numberwithin{equation}{section}
\numberwithin{thm}{section}

\newcommand{\gen}[1]{\partial_{#1}}
\newcommand{\pr}[1]{\rm pr^{(#1)}}
\newcommand{\pc}{\psi^\ast}
\newcommand{\curl}[1]{ \{#1\} }

\newcommand{\R}{\mathbb{R}}

\newcommand{\semi}{\subset \hskip -3.8mm +}
\newcommand{\semiprod}{\subset \hskip -3.8mm \times}
\newcommand{\sch}{\mathfrak {sch}}
\newcommand{\heis}{\mathfrak h}

\DeclareMathOperator{\so}{so} \DeclareMathOperator{\Sl}{sl}

 \DeclareMathOperator{\cn}{cn}

\begin{document}

\title{\bf
\Large Group-invariant Solutions of the $2+1$-dimensional Cubic
Schr\"{o}dinger Equation}

\author{
C. \"{O}zemir\thanks{e-mail: ozemir@itu.edu.tr} and F.~G\"ung\"{o}r\thanks{e-mail: gungorf@itu.edu.tr}\\
\small Department of Mathematics, Faculty of Science and Letters,\\
\small Istanbul Technical University, 34469 Istanbul, Turkey}

\date{}

\maketitle

\begin{abstract}
We use  Lie point symmetries of the 2+1-dimensional cubic
Schr\"{o}dinger equation to obtain new analytic solutions in a
systematic manner. We present an analysis of the reduced ODEs, and
in particular show that although the original equation is not
integrable they typically can belong to the  class of Painlev\'e
type equations.

\end{abstract}

\section{Introduction}

The purpose of this article is to find new explicit group
invariant solutions of the cubic Schr\"{o}dinger equation (CSE)
\begin{equation}\label{cse}
    i\psi_t+\Delta \psi=a_0|\psi|^2\psi
\end{equation}
where $\psi(t,x,y)$ is a complex-valued function of its arguments,
$a_0$ is a real constant and $\Delta$ is the Laplace operator in
$\mathbb{R}^2$.

Eq. \eqref{cse} plays an important role in many branches of
physics. For instance, in nonlinear optics it describes the
interaction of electromagnetic waves with different polarizations
propagating in nonlinear media such as an isotropic plasma.
Another instance where Eq. \eqref{cse} arises is water waves in a
deep fluid.

In one space dimension, the CSE is known to be integrable via
inverse scattering transformation. In two space dimensions this
approach does not work. Therefore, our approach will  rely on the
use of powerful methods of symmetries of differential equations.
The Lie group $G$ of local point transformations leaving the
equation invariant is well-known in literature. We shall use
subgroups of $G$ to reduce the equation to a system of algebraic
or (second order) ordinary differential equations. At this stage,
it is of vital importance to perform a classification of
subalgebras. The reduced equations will be compatible with
specific types of boundary conditions, such as translational
invariance or a cylindrical symmetry. Then, they will be solved
whenever possible, in particular when they have the Painlev\'e
property. This approach is basically similar to that of a
generalized non-linear Schr\"{o}dinger equation involving cubic,
quintic equations as special cases in three space dimensions
($n=3$) studied  in a series of papers \cite{Gagnon88, Gagnon89-2,
Gagnon89-3}. For a vector version of \eqref{cse}, a similar
analysis for two space dimensions ($n=2$) and two or three wave
functions ($N=2,3$) is given in Ref. \cite{Sciarrino97}. The
present article will complete the results of \cite{Sciarrino97} to
include the scalar case, namely one wave ($N=1$).

We organize the article as follows. In Section 2, we review the
symmetry group $G$,  its algebra $L$ and subalgebras of the
Schr\"{o}dinger equation. In Section 3, we utilize subalgebras to
perform reductions to second order  ordinary differential
equations (ODEs).  In Section 4 we construct solutions of the
reduced equations whenever possible and thereby invariant
solutions of the original equation. In particular, we show that
for specific values of group parameters, we can pick out equations
having Painlev\'e property among generically non-Painlev\'e ones.
Finally, in Section 5 we make some concluding remarks.

\section{The symmetry group and its Lie algebra}

By a symmetry group $G$ we shall refer to the local Lie group of
point transformations
\begin{equation}
\begin{split}
    \tilde{t}&=T_g(t,x,y,\psi,{\psi}^{*}), \quad
    \tilde{x}=X_g(t,x,y,\psi,{\psi}^{*}),\quad
    \tilde{y}=Y_g(t,x,y,\psi,{\psi}^{*}),\\
    \tilde{\psi}&=\Psi_g(t,x,y,\psi,{\psi}^{*})
\end{split}
\end{equation}
such that $\tilde{\psi}(t,x,y)$ is a solution whenever
$\psi(t,x,y)$ is one. Here star denotes complex conjugation and
$g$ is the group element.

The computation of the symmetry group is quite straightforward
(for details see for example \cite{Olver91}). To this end, we look
for the infinitesimal symmetries that close to form a Lie algebra
under commutation. The Lie algebra $L$ of the symmetry group $G$
is realized by vector fields of the form
\begin{equation}\label{vf}
    \mathbf{v}=\tau\gen t+\xi\gen x+\eta\gen y+\varphi\gen \psi+{\varphi^{*}}\gen
    {{\psi}^{*}}.
\end{equation}
The coefficients $\tau$, $\xi$, $\eta$, $\varphi$, $\varphi^{*}$
are functions of $t,x,y,\psi,{\psi}^{*}$ to be determined from the
 invariance condition
\begin{equation}\label{symcon}
    {\pr{2}}\mathbf{v}(E)\Bigl|_{E=0}=0,
\end{equation}
where $E$ is the equation under study and $\pr{2}$ is the second
prolongation of the vector field $\mathbf{v}$. This condition
provides us a system of determining equations for the coefficients
of $\mathbf{v}$. Solving this system we end up with a basis
(depending on integration constants) of vector fields for the
symmetry algebra $L$. We exponentiate them, namely calculate $\exp
L$ to find the symmetry group $G$.

An inverse problem of determining the most general form of
Schr\"{o}dinger type equations
\begin{equation}\label{genSE}
\psi_t+F(x,y,t,\psi_i,\pc_i,\psi_{ij},\pc_{ij})=0,\quad
i,j\in\curl{x,y}
\end{equation}
invariant under the Schr\"{o}dinger and its subgroups was attacked in
\cite{Gungor99}. An investigation of this problem was reduced to
that of classifying inequivalent realizations of the Schr\"{o}dinger
algebra. In particular, the CSE appears to be among those
equations allowing  the same Schr\"{o}dinger algebra $\sch(2)$ as its
symmetry algebra. Hence, we simply resort to the results of
\cite{Gungor99} for recovering a suitable basis of the algebra
written in phase and modulus coordinates. In passing, we mention
that symmetry breaking interactions for the time dependent
Schr\"{o}dinger equation were studied in \cite{Boyer76} for $n=1$ and
in \cite{Gungor00} for $n=2$.

We introduce the moduli and phases of $\psi$ by setting
\begin{equation}
    \psi=Re^{i\phi},\quad 0\leq R<\infty,\quad 0\leq \phi\leq
    2\pi.
\end{equation}
and rewrite our equation \eqref{cse} in terms of $R$ and $\phi$ as
a coupled system of real equations
\begin{subequations}\label{sys}
\begin{eqnarray}
      -R\phi_t+\Delta R-R|\nabla \phi|^2 &=&a_0 R^3, \\
      R_t+R\Delta \phi+2\nabla R\cdot \nabla \phi &=&0 \label{sysb},
\end{eqnarray}
\end{subequations}
where $R$ and $\phi$ are functions of $t, x, y$.

We summarize some facts on the structure of the symmetry algebra
and its important subalgebras.
\begin{itemize}
    \item The nine-dimensional Schr\"{o}dinger algebra $\sch(2)$ has a Levi
decomposition
\begin{align}\label{sch}
\sch(2)&=\heis_2 \semi\curl{\Sl(2,\R)\oplus\so(2)} \nonumber\\
&\sim\curl{P_1, P_2, B_1, B_2, E}\semi\curl{T,C,D,J}
\end{align}
with nonzero commutation relations
\begin{align}\label{comm}
&[P_1,B_1]=E/2,  &  [P_2,B_2]&=E/2,\nonumber\\
&[J,B_2]=-B_1, & [J,P_2]&=-P_1, \nonumber\\
&[J,B_1]=B_2, & [J,P_1]&=P_2, \nonumber\\
&[T,B_j]=P_j, & [D,B_j]&=B_j, \quad j=1,2\\
&[D,P_j]=-P_j, & [C,P_j]&=-B_j,  \quad j=1,2\nonumber\\
&[T,D]=2T,     & [T,C]&=D, \quad [D,C]=2C. & \nonumber
\end{align}
Here $\semi$ denotes a semi-direct sum and $\heis_2$  is a
Heisenberg algebra.

    \item The subalgebra $\curl{T, P_1, P_2, B_1, B_2, J, E}$ generates the
extended Galilei group (translations, proper Galilei
transformations and constant change of phase), $D$ and $C$
generate dilations and nonrelativistic conformal transformations.
    \item A convenient basis for the standard Schr\"{o}dinger algebra is given
by the vector fields
\begin{subequations}\label{gen}
\begin{eqnarray}
    &&T=\gen t,\quad P_1=\gen x,\quad P_2=\gen y,\quad E=\gen
    \phi,\\
    &&B_1=t\gen x+x/2 \gen \phi,\quad B_2=t\gen y+y/2 \gen \phi, \quad J=-y\gen x+x\gen y,\\
    &&D=2t\gen t+x\gen x+y\gen y-R\gen R,\\
    &&C=t^2\gen t+xt\gen
    x+yt\gen y-tR\gen R+\frac{1}{4}(x^2+y^2)\gen \phi.
\end{eqnarray}
\end{subequations}

\item
The transformations corresponding to the Galilei-similitude
algebra are global and given by
\begin{equation}\label{global}
\begin{split}
    \tilde{t}&=e^{\lambda}(t-t_0),\\
    \tilde{x}&=e^{\lambda/2}\curl{
    \cos \alpha[(x-x_0)+v_1(t-t_0)]+\sin
    \alpha[(y-y_0)+v_2(t-t_0)]},\\
\tilde{y}&=e^{\lambda/2}\curl{
    -\sin \alpha[(x-x_0)+v_1(t-t_0)]+\cos
    \alpha[(y-y_0)+v_2(t-t_0)]},\\
    \tilde{R}&=e^{-\lambda/2}R, \quad
    \tilde{\phi}=\phi+\frac{1}{2}[v_1(x-x_0)+v_2(y-y_0)]+
    \frac{1}{4}(v_1^2+v_2^2)(t-t_0)+\delta,
\end{split}
\end{equation}
where $t_0, x_0, y_0, v_1, v_2, \alpha, \lambda, \delta$ are group
parameters.

\item The conformal transformations generated by $C$ are local and given
by
\begin{equation}\label{conf}
\begin{split}
    \tilde{t}&=\frac{t}{1-pt},\quad \tilde{x}=\frac{x}{1-pt},\quad
    \tilde{y}=\frac{y}{1-pt},\\
    \tilde{R}&=(1-pt)R,\quad
    \tilde{\phi}=\phi+\frac{p(x^2+y^2)}{4(1-pt)},
\end{split}
\end{equation}
provided $1-pt\ne 0$. Above, $p$ is  group parameter.

The transformations \eqref{global} and \eqref{conf} can be
composed to obtain the transformations of the Schr\"{o}dinger group.

\item The CSE is also invariant under time reversal and reflection
of the space coordinates:
\begin{equation}\label{discrete}
\begin{split}
T_d&: \quad t\to -t,\quad x\to x,\quad y\to y,\quad \psi\to
{\psi}^{*},\\
R_x&: \quad t\to t,\quad x\to -x,\quad y\to y,\quad \psi\to
{\psi},\\
R_y&: \quad t\to t,\quad x\to x,\quad y\to -y,\quad \psi\to
{\psi}.
\end{split}
\end{equation}
The elements $T, R_x, R_y$ generate a discrete finite group $G_D$.
\end{itemize}

\noindent{\bf Comment:}  The conformal transformations generated
by $C$ do not generalize to higher dimensional cubic Schr\"{o}dinger
equations. We mention that in a higher dimensional space $n$, the
nonlinear Schr\"{o}dinger equation with a critical exponent $4/n$
\cite{Fushchich93}
$$i\psi_t+\Delta\psi=|\psi|^{4/n}\psi,\quad x\in \mathbb{R}^n$$
is conformally invariant. In this case, the conformal generator
$C$ and the corresponding local group action on $(x,t,R,\phi)$ are
given by
\begin{equation}
    C=t^2\gen t+t\sum_{j=1}^{n}x_j\gen {x_j}-\frac{nt}{2}R\gen
R+\frac{|x|^2}{4}\gen \phi,\quad |x|^2=\sum_{j=1}^{n} x_j^2,
\end{equation}
\begin{equation}\label{confn}
\begin{split}
    \tilde{t}&=\frac{t}{1-pt},\quad
    \tilde{x}_j=\frac{x_j}{1-pt},\quad j=1,\ldots,n\\
    \tilde{R}&=(1-pt)^{n/2}R,\quad
    \tilde{\phi}=\phi+\frac{p|x|^2}{4(1-pt)},
\end{split}
\end{equation}
provided $1-pt\ne 0$. The induced action of $C$ on a solution
$\psi_0(x,t)$ produces the transformation formula
$$\psi(x,t)=(1+pt)^{-n/2}\exp\Bigl(\frac{ip|x|^2}{4(1+pt)}\Bigr)
\psi_0\Bigl(\frac{x}{1+pt},\frac{t}{1+pt}\Bigr),\quad x\in
\mathbb{R}^n$$ which means that $\psi(x,t)$ is a solution whenever
$\psi_0(x,t)$ is one.

\subsection{Subalgebras of the Schr\"{o}dinger algebra $\sch(2)$}

To perform symmetry reduction for Eq. \eqref{cse} we need a
classification of low-dimensional subalgebras of the extended
Schr\"{o}dinger  algebra $L=\sch(2)$ into conjugacy classes under the
action of the  group of transformations $G$ leaving equation
\eqref{cse} invariant. The group $G$ has the structure
$$G=G_D\semiprod G_0,$$
where $\semiprod$ denotes a semi-direct product.  The invariant
subgroup $G_0$ is the connected component of $G$, the Lie algebra of
which is \eqref{sch} and $G_D$ is the discrete component defined by
\eqref{discrete}.
 In order to reduce Eq. \eqref{cse} to an algebraic
equation,  an ordinary differential equation (ODE) or a partial
differential equation (PDE) in two variables we need to know all
subalgebras $L_0\subset L$ of dimension $\dim L_0=3,2$ and 1,
respectively. Moreover, the corresponding subalgebras should not
contain the element $E$ alone as long as we are interested in
invariant solutions. Otherwise, the subgroup invariants would be
independent of the phase $\phi$ and this in turn would not define
locally invertible transformations to $\psi$ or $\psi^*$.

A complete list of representatives of all subalgebra classes of
$\sch(2)$ are given in Ref. \cite{Burdet78-1}.

\section{Symmetry Reductions}
A group-invariant solution is one that is transformed to itself by
group transformations. To find such a solution we choose a
suitable canonical subgroup and require that the solution be left
invariant. In this section we are interested in looking at
solutions invariant under two-dimensional subalgebras since they
lead to reductions to real systems of ODEs. We go through each
representative subalgebra and find the corresponding reduced
system and solve them whenever possible, and thereby a solution of
the original PDE. One can transform these solutions by group
transformations to obtain all other possible invariant solutions.
Note that the ideas we employ here are described more fully in
\cite{Gagnon88, Gagnon89-2, Gagnon89-3}.

In table \ref{Tab1} we list representatives of two-dimensional
subalgebras that provide reductions to ODEs. We note that we
skipped those trivially acting on the coordinate space $(t, x,
y)$.

\begin{table}\caption{Two-dimensional subalgebras leading to ODEs ($\varepsilon=\pm 1$)}\label{Tab1}
\begin{center}
\begin{tabular}{|c|c|c|}
\hline
    No    & Type & Basis \\
\hline
 $L_{2,1}$ & 2$A_1$     & $J+aE, C+T+bE$      \\
 $L_{2,2}$         &      & $J+aE, T$      \\
 $L_{2,3}$         &      & $J+aE, T+\varepsilon E$      \\
 $L_{2,4}$         &      & $J+aE, D+b E$      \\
 $L_{2,5}$         &      & $T, P_1$      \\
 $L_{2,6}$         &      & $T+\varepsilon E, P_1$      \\
 $L_{2,7}$         &      & $T+B_1, P_2$      \\
 $L_{2,8}$         &      & $P_1, P_2$      \\
 $L_{2,9}$         &      & $B_1, P_2$      \\
 $L_{2,10}$         &      &$J-\varepsilon(C+T)+aE, B_1+\varepsilon P_2$       \\
\hline
 $L_{2,11}$         &  $A_2$    & $D+aJ+bE, T$       \\
 $L_{2,12}$         &      &  $D+aE, P_1$     \\
\hline
\end{tabular}
\end{center}
\end{table}

Subgroups that have generic orbits of codimension one in the
coordinate space $(t, x, y)$ and of codimension  three in the
total space $(t, x, y, R, \phi)$ will lead to reductions to ODEs.

The procedure for finding invariant solutions is to require that the solution $\psi(t,x,y)$ be invariant
under each subalgebra. This  invariance imposes constraints on the solution that are expressed
by  first order linear PDEs. Their solutions imply that the invariant solutions should have the form
\begin{equation}\label{ansatz}
    \psi(t,x,y)=f(t,x,y)M(\xi)\exp{[i(g(t,x,y)+\Phi(\xi))]},\quad M(\xi),
    \Phi(\xi)\in \mathbb{R},
\end{equation}
where $f(t,x,y)$, $g(t,x,y)$ and $\xi(t,x,y)$ (the subgroup
invariant) are explicitly known for each subalgebra. Below we give
the expressions for $\xi$ and the wave $\psi(t,x,y)$ together with
the reduced equations for the functions $M(\xi)$ and $\Phi(\xi)$ in
a list. Note that for all of the abelian subalgebras, one  of the
coupled reduced equations, namely reduction of \eqref{sysb} is
directly integrable once. Using this we eliminate $\Phi(\xi)$ to
obtain a second order real ODE for $M(\xi)$. The non-abelian cases
are exceptionally difficult to handle. In the last case
($L_{2,12}$), we raise the order of the ODE to make a decoupling
possible and hence have to deal with a third order ODE. The
subalgebra $L_{2,11}$ is the only case that has led to a reduction
which we have not been able to succeed in treating analytically in
any way.

\begin{itemize}
\item{\bf The subalgebra $L_{2,1}$}
\begin{equation}\label{21}
\begin{split}
    &\psi(t,x,y)=\frac{M(\xi)}{\sqrt{1+t^2}}e^{i(a\theta+b\arctan
    t+t\xi/4+\Phi(\xi))},\\
    &\xi=\frac{x^2+y^2}{1+t^2},\quad \theta=\arctan\frac{y}{x},
\end{split}
\end{equation}
\begin{subequations}
\begin{eqnarray}
   \label{21a}  &&M^2\dot{\Phi}\xi=C_0,\quad C_0=\text{const.} \\
   \label{21b}  &&\ddot{M}=(\frac{b}{4\xi}+\frac{a^2}{4\xi^2}+\frac{1}{16})M-
                \frac{\dot{M}}{\xi}+\frac{C_0^2}{\xi^2}M^{-3}+\frac{a_0}{4\xi}M^3.
\end{eqnarray}
\end{subequations}

\item{\bf The subalgebra $L_{2,2}$}
\begin{equation}\label{22}
\begin{split}
    &\psi(t,x,y)=M(\rho)e^{i(a\theta+\Phi(\rho))},\\
    &\rho=(x^2+y^2)^{1/2},\quad \theta=\arctan\frac{y}{x},
\end{split}
\end{equation}
\begin{subequations}\label{22red}
\begin{eqnarray}
   \label{22a}  &&M^2\dot{\Phi}\rho=C_0,\quad C_0=\text{const.} \\
   \label{22b}  &&\ddot{M}=\frac{a^2}{\rho^2}M+\frac{C_0^2}{\rho^2}M^{-3}-
                 \frac{\dot{M}}{\rho}+a_0M^3.
\end{eqnarray}
\end{subequations}

\item{\bf The subalgebra $L_{2,3}$}
\begin{equation}\label{23}
\begin{split}
    &\psi(t,x,y)=M(\rho)e^{i(a\theta+\varepsilon\ln t+\Phi(\rho))}, \\
    &\rho=(x^2+y^2)^{1/2},\quad \theta=\arctan\frac{y}{x},
\end{split}
\end{equation}
\begin{subequations}
\begin{eqnarray}
  \label{23a} &&M^2\dot{\Phi}\rho=C_0,\quad C_0=\text{const.} \\
  \label{23b} &&\ddot{M}=(\varepsilon+\frac{a^2}{\rho^2})M+\frac{C_0^2}{\rho^2}M^{-3}-
              \frac{\dot{M}}{\rho}+a_{0}M^3.
\end{eqnarray}
\end{subequations}

\item{\bf The subalgebra $L_{2,4}$}
\begin{equation}\label{24}
\begin{split}
    &\psi(t,x,y)=\frac{M(\xi)}{\sqrt{t}}e^{i(a\theta+\frac{b}{2}\ln t+\Phi(\xi))},\\
    &\xi=\frac{x^2+y^2}{t},\quad \theta=\arctan\frac{y}{x},
\end{split}
\end{equation}
\begin{subequations}
\begin{eqnarray}
   \label{24a}  &&M^2(8\dot{\Phi}-1)\xi=C_0,\quad C_0=\text{const.} \\
   \label{24b}  &&\ddot{M}=(\frac{b}{8\xi}+\frac{a^2}{4\xi^2}-\frac{1}{64})M+\frac{C_0^2}{64\xi^2}M^{-3}
                 +\frac{a_{0}}{4\xi}M^3-\frac{\dot{M}}{\xi}.
\end{eqnarray}
\end{subequations}

\item{\bf The subalgebra $L_{2,5}$}
\begin{equation}\label{25}
\begin{split}
    &\psi(t,x,y)=M(y)e^{i\Phi(y)},\\
\end{split}
\end{equation}
\begin{subequations}
\begin{eqnarray}
   \label{25a}  &&M^2\dot{\Phi}=C_0,\quad C_0=\text{const.}    \\
   \label{25b}  &&\ddot{M}=C_0^2M^{-3}+a_{0}M^3.
\end{eqnarray}
\end{subequations}

\item{\bf The subalgebra $L_{2,6}$}
\begin{equation}\label{26}
\begin{split}
    &\psi(t,x,y)=M(y)e^{i(\varepsilon t+\Phi(y))},\\
\end{split}
\end{equation}
\begin{subequations}
\begin{eqnarray}
    \label{26a}  &&M^2\dot{\Phi}=C_0,\quad C_0=\text{const.}      \\
    \label{26b}  &&\ddot{M}=\varepsilon M+C_0^2M^{-3}+a_{0}M^3.
\end{eqnarray}
\end{subequations}

\item{\bf The subalgebra $L_{2,7}$}
\begin{equation}\label{27}
\begin{split}
    &\psi(t,x,y)=M(\xi)e^{i(\frac{xt}{2}-\frac{t^{3}}{6}+\Phi(\xi))},\\
    &\xi=x-\frac{t^{2}}{2},
\end{split}
\end{equation}
\begin{subequations}
\begin{eqnarray}
   \label{27a}  &&M^2\dot{\Phi}=C_0,\quad C_0=\text{const.} \\
   \label{27b}  &&\ddot{M}=\frac{\xi}{2}M+C_0^2M^{-3}+a_{0}M^3.
\end{eqnarray}
\end{subequations}

\item{\bf The subalgebra $L_{2,8}$}
\begin{equation}\label{28}
\begin{split}
    &\psi(t,x,y)=M(t)e^{i\Phi(t)},\\
\end{split}
\end{equation}
\begin{subequations}
\begin{eqnarray}
   \label{28a}  &&\Phi=-aC_0^{2}t+C_{1},\quad C_0,C_1=\text{const.} \\
   \label{28b}  &&M=C_0.
\end{eqnarray}
\end{subequations}

\item{\bf The subalgebra $L_{2,9}$}
\begin{equation}\label{29}
\begin{split}
    &\psi(t,x,y)=M(t)e^{i(\frac{x^{2}}{4t}+\Phi(t))},\\
\end{split}
\end{equation}
\begin{subequations}
\begin{eqnarray}
    \label{29a} &&M=C_0t^{-\frac{1}{2}}, \\
    \label{29b} &&\Phi=-2aC_0t^{\frac{1}{2}}+C_1, \quad C_0, C_1=\text{const.}
\end{eqnarray}
\end{subequations}

\item{\bf The subalgebra $L_{2,10}$}
\begin{equation}\label{30}
\begin{split}
   &\psi(t,x,y)=\frac{M(\xi)}{\sqrt{1+t^{2}}}e^{i\phi(t,x,y)},\\
   &\phi(t,x,y)=\frac{x^{2}}{4t}-a\varepsilon \arctan
   t+\frac{\xi^{2}(t^{2}-1)}{4t}+\Phi(\xi),\\
   &\xi=\frac{\varepsilon x-yt}{1+t^{2}},
\end{split}
\end{equation}
\begin{subequations}
\begin{eqnarray}
   \label{30a}  &&M^2\dot{\Phi}=C_0,\quad C_0=\text{const.} \\
   \label{30b}  &&\ddot{M}=(\xi^{2}-a\varepsilon) M+C_0^2M^{-3}+a_{0}M^3.
\end{eqnarray}
\end{subequations}

\item{\bf The subalgebra $L_{2,11}$}
\begin{equation}\label{31}
\begin{split}
    &\psi(t,x,y)=\frac{M(\xi)}{\rho}e^{i\bigl(\frac{b}{a}\theta+\Phi(\xi)\bigr)},\\
    &\xi=\rho\exp(-\theta/a),\\
    &\rho=(x^2+y^2)^{1/2},\quad \theta=\arctan\frac{y}{x},
\end{split}
\end{equation}
\begin{equation}
\begin{split}
    &(1+a^2)\xi M\ddot{\Phi}+2(1+a^2)\xi\dot{M} \dot{\Phi}+(1-a^2)M\dot{\Phi}-2b\dot{M}=0,\\[.2cm]
    &(a^2-b^2)M+(1-a^2)\xi\dot{M}+(1+a^2)\xi^2\ddot{M}+2b\xi M\dot{\Phi}-(1+a^2)\xi^2M\dot{\Phi}^2\\
   &=a^2 a_0 M^3.
\end{split}
\end{equation}

\item{\bf The subalgebra $L_{2,12}$}
\begin{equation}\label{32}
\begin{split}
    &\psi(t,x,y)=\frac{M(\xi)}{\sqrt{t}}e^{i\bigl(\frac{a}{2}\ln t+\Phi(\xi)\bigr)},\\
    &\xi=\frac{y}{\sqrt{t}}, \\
\end{split}
\end{equation}

\begin{subequations}
\begin{eqnarray}
   \label{32a}&&\dot{\Phi}=\frac{C_0+\xi+\xi \dot{\chi}}{4\dot{\chi}} , \quad \dot{\chi}=M^2, \quad C_0=\text{const.} \\
   \label{32b}&&2\dot{\chi} \dddot{\chi}-\ddot{\chi}^2+(\frac{\xi^2}{4}-2a)\dot{\chi}^2-\frac{1}{4}\chi^2-4a_0\dot{\chi}^3
                -\frac{C_0}{2}\chi-\frac{C_0^2}{4}=0.
\end{eqnarray}
\end{subequations}

\end{itemize}

\section{Discussion of the solutions of reduced ODEs}
In Section 3,  we  showed that for the abelian subalgebras of type
$2A_1$ the  reduced pair of equations can always be decoupled by
integrating the second one once. For the nonabelian case of type
$A_2$, decoupling is possible only for one subalgebra at the price
of introducing third order derivative. Luckily, it happens to have
the Painlev\'e property and its solution is expressible in terms
of the fourth Painlev\'e transcendent $P_{IV}$.

For each reduction corresponding to subalgebras $L_{2,i}$,
$i=1,\ldots ,10$, $M(\xi)$ satisfies a second order ordinary
differential equation. All of them can be written in a unified
form (see Table \ref{Tab2})
\begin{equation}\label{single}
    \ddot{M}=B_1(\xi)\dot{M}+A_1(\xi)M+A_3(\xi)M^3+B_0(\xi)M^{-3}.
\end{equation}
We change the dependent variable to $H$ by putting
\begin{equation}\label{posH}
    M=\sqrt{H},\quad H>0
\end{equation}
so that the Eq. is transformed to
\begin{equation}\label{single1}
    \ddot{H}=\frac{1}{2H}\dot{H}^2+B_1\dot{H}+2A_1H+2A_3H^2+\frac{2B_0}{H}.
\end{equation}
To simplify \eqref{single1} we apply the transformation
\begin{equation}\label{trans}
    H=\lambda(\xi)W(\eta),\quad \eta=\eta(\xi)
\end{equation}
and obtain
\begin{equation}\label{single2}
\begin{split}
    \ddot{W}&=\frac{1}{2W}\dot{W}^2+\left(\frac{B_1}{\dot{\eta}}-\frac{\ddot{\eta}}{\dot{\eta}^2}
    -\frac{1}{\dot{\eta}}\frac{\dot{\lambda}}{\lambda}\right)\dot{W}+
    (\frac{2A_1}{\dot{\eta}^2}+\frac{\dot{\lambda}}{\lambda}\frac{B_1}{\dot{\eta}^2}+
    \frac{\dot{\lambda}^2}{2\lambda^2\dot{\eta}^2}-\frac{\ddot{\lambda}}{\lambda\dot{\eta}^2})W\\
    &+\frac{2\lambda A_3}{\dot{\eta}^2}W^2+\frac{2B_0}{\lambda^2\dot{\eta}^2}W^{-1}.
\end{split}
\end{equation}

Eq. \eqref{single2} can be transformed to one of the canonical
equations classified by  Painlev\'e and Gambier (see \cite{Ince56}
for a list of canonical equations). We shall see that only for
special values of coefficients  they can be integrated in terms of
solutions of linear equations, elliptic functions or Painlev\'e
transcendents. Otherwise, they would not have Painlev\'e property
which means that they can not be integrated.

\begin{table}\caption{Coefficients of \eqref{single}}\label{Tab2}
\begin{center}
\begin{tabular}{|c|c|c|c|c|}
\hline
    No      & $A_1$                       & $A_3$       & $B_0$             &$B_1$\\
\hline
 $L_{2,1}$  &$b/(4\xi)+a^2/(4\xi^2)+1/16$ &$a_0/(4\xi)$ & $C_0^2/\xi^2$     & $-1/\xi$    \\
 $L_{2,2}$  &$a^2/\rho^2$                 &$a_0$        & $C_0^2/\rho^2$    & $-1/\rho$    \\
 $L_{2,3}$  &$\varepsilon+a^2/\rho^2$     &$a_0$        & $C_0^2/\rho^2$    & $-1/\rho$     \\
 $L_{2,4}$  &$b/(8\xi)+a^2/(4\xi^2)-1/64$ &$a_0/(4\xi)$ & $C_0^2/(64\xi^2)$ & $-1/\xi$    \\
 $L_{2,5}$  &$0$                          &$a_0$        & $C_0^2$           & $0$     \\
 $L_{2,6}$  &$\varepsilon$                &$a_0$        & $C_0^2$           & $0$    \\
 $L_{2,7}$  &${\xi}/{2}$              &$a_0$        & $C_0^2$           & $0$    \\
 $L_{2,10}$ &$\xi^2-a\varepsilon$         &$a_0$        & $C_0^2$           & $0$  \\
\hline
\end{tabular}
\end{center}
\end{table}

\subsection{Solutions using the subalgebra $L_{2,1}$}
Eq. \eqref{single2} for \eqref{21b} is
\begin{equation}\label{211}
\begin{split}
    \ddot{W}&=\frac{1}{2W}\dot{W}^2-\frac{1}{\dot\eta}(\frac{1}{\xi}+\frac{\ddot{\eta}}{\dot{\eta}}
    +\frac{\dot{\lambda}}{\lambda})\dot{W}
    +\frac{a_0\lambda}{2\xi\dot{\eta}^2}W^2+\frac{2C_0^2}{(\xi\lambda\dot{\eta)}^2}W^{-1}\\
    &+\frac{1}{\dot{\eta}^2}(\frac{b}{2\xi}+\frac{a^2}{2\xi^2}+\frac{1}{8}-\frac{\dot{\lambda}}{\lambda \xi}+
    \frac{\dot{\lambda}^2}{2\lambda^2}-\frac{\ddot{\lambda}}{\lambda})W.
\end{split}
\end{equation}
This equation may be transformed to either PXXXIII or PXXXIV among
the canonical equations (\cite{Ince56, Davis62}). For
completeness,  we quote them here:
\begin{eqnarray}
\label{P33} &&\rm{PXXXIII:} \qquad \ddot{W}=\frac{1}{2W}\dot{W}^2+4W^2+\gamma W-\frac{1}{2}W^{-1}, \\
\label{P34} &&\rm{PXXXIV:}  \qquad
\ddot{W}=\frac{1}{2W}\dot{W}^2+4\gamma W^2-\eta
W-\frac{1}{2}W^{-1}.
\end{eqnarray}
Comparing the coefficient of $W^2$ and $\dot{W}$ in \eqref{211}
with  \eqref{P33} and \eqref{P34} we see that $\lambda$ and $\eta$
must be chosen as
\begin{equation}\label{lami}
\lambda=\lambda_0\xi^{-1/3}, \quad \eta=\eta_0\xi^{1/3}.
\end{equation}
For this choice of $\lambda$ and $\eta$ \eqref{211} becomes
\begin{equation}\label{212}
\ddot{W}=\frac{1}{2W}\dot{W}^2+\frac{9a_0\lambda_0}{2\eta_0^2}W^2+18(\frac{C_0}{\lambda_0\eta_0})^2W^{-1}
+\left[\frac{(9a^2-1)}{2}\eta^{-2}+\frac{9b}{2\eta_0^3}\eta+\frac{9}{8\eta_0^6}\eta^4\right]W.
\end{equation}
Inspecting \eqref{P33} and \eqref{P34}, we see that the
coefficient of $W$ must be a constant or $\eta$. For Eq.
\eqref{212}, in the coefficient of $W$,  it is possible to remove
the term $\eta^{-2}$ by choosing $a^2=1/9$ but we cannot get rid
of the term $\eta^4$. Hence Eq. \eqref{21b} cannot be transformed
to one of the canonical forms, thus it does not have the
Painlev\'e property.

If we consider Eq.\eqref{21b} with $C_0=0$, similar arguments
apply. We mention that Eq. \eqref{21b} admits no point symmetries
and no constant solutions.

\subsection{Solutions using the subalgebra $L_{2,2}$}

\subsubsection{Equation \eqref{22b}, with $C_{0}\neq0$} By a suitable
choice of $\lambda$ and $\eta$ this equation can be transformed to
a standard form  which has no first derivative and a constant
coefficient of the quadratic term. The transformation and the
corresponding transformed equation are given by

$$\lambda=\lambda_0\rho^{-2/3},\quad
\eta=\eta_0\rho^{2/3},  $$
$$\ddot{W}=\frac{1}{2W}\dot{W}^2+(\frac{9a_0\lambda_0}{2\eta_0^2})W^2
+(\frac{9a^2-1}{2}\eta^{-2})W+
\frac{9}{2}(\frac{C_0}{\lambda_0\eta_0})^{2}W^{-1}.$$

The above equation has the Painlev\'e property
only when $a^2=1/9$. In this case we set
$\eta_0^{2}=9a_0\lambda_0/8$ and integrate once to obtain
\begin{equation}\label{2,33}
\dot{W}^2=4(W^3+CW-\delta^{2}), \quad C=\text{const.} , \quad \delta^2=\frac{2C_0^2}{a_0\lambda_0^3}\\
\end{equation}
whose solution can be expressed in terms of Weierstrass $\wp(\xi)$
function or Jacobi elliptic functions \cite{Byrd71}. By virtue of
\eqref{posH} we require $\lambda W$ to be positive.  Writing the
right-hand side of \eqref{2,33} as
$$P(W)=4(W^3+CW-\delta^{2})=4(W-W_1)(W-W_2)(W-W_3),$$
where  $W_1, W_2, W_3$ are roots of the cubic polynomial $P(W)$,
we investigate the solutions of \eqref{2,33} for four different
cases:

(i) $P(W)$ has only one root. This is not possible since $P(W)$
does not include the term $W^{2}$.

(ii) $P(W)$ has a double root, say $W_{1}=W_{2}\neq W_{3}$ and
write $P(W)$ as
$$P(W)=4(W^3+CW-\delta^{2})=4(W-W_1)^{2}(W-W_3).$$
This case is possible for
$C=-3.2^{\frac{-2}{3}}\delta^{\frac{4}{3}}$ and $W_{1,2,3}$ are
given as $W_{1,2}=-2^{-\frac{1}{3}}\delta^{\frac{2}{3}}$,
$W_{3}=(2\delta)^{\frac{2}{3}}$. Integrating for  $W>W_{3}$ we
reach to the solution given below:
\begin{equation}
\begin{split}
 &M=\frac{2 c_{1}}{3}(\frac{2}{a_{0}})^{\frac{1}{2}}\rho^{-\frac{1}{3}}(-\frac{1}{3}+\sec^{2}\tau)^{\frac{1}{2}},
 \qquad \tau=c_{1}\rho^{\frac{2}{3}}+c_{2}, \qquad
 c_{1}^{2}=\frac{27}{8}(a_{0}C_{0})^{\frac{2}{3}}\\
 &\Phi=\tan^{-1}(\sqrt{\frac{3}{2}}\tan\tau)-\sqrt{\frac{2}{3}}\tau+\Phi_{0},
 \qquad \psi=M\exp[i(\pm\frac{1}{3}\theta+\Phi)].\\
\end{split}
\end{equation}

(iii)  $P(W)$  has three distinct roots. In this case, roots
$W_{1,2,3}$ are determined by the system
$$W_1+W_2+W_3=0,\quad W_1W_2+W_1W_3+W_2W_3=C,\quad
W_1W_2W_3=\delta^{2}.$$ \\ We see that two of the roots have the
same sign, while the third is always positive. Their solutions
depend on ordering of the roots.
$$(a)\quad W_{1}<W_{2}<0<W_{3} \quad (b) \quad 0<W_{1}<W_{2}<W_{3}, \quad
W_{2}^{2}<\frac{\delta^{2}}{W_{1}}.$$
 Integration for $W>W_{3}$ gives both for (a) and (b)
\begin{eqnarray}
\begin{split}
  &W=W_{2}+(W_{3}-W_{2})[\textrm{cn}(\tau,k)]^{-2}, \quad &&\tau=c_{1}\rho^{\frac{2}{3}}+c_{2}, \quad k^{2}=\frac{W_{2}-W_{1}}{W_{3}-W_{1}},\\
  &M= \rho^{-1/3}(\lambda_{0}W)^{1/2},
  \qquad &&c_{1}^{2}=\frac{9a_{0}\lambda_{0}}{8}(W_{3}-W_{1}), \\
  &\Phi=\frac{\delta}{\sqrt{W_{3}-W_{1}}}\int\frac{d\tau}{W}+\Phi_{0},
   \qquad &&\psi=M\exp[i(\pm\frac{1}{3}\theta+\Phi)]\\
\end{split}
\end{eqnarray}
with $a_{0},\lambda_{0}>0$. Integrating on $W_{1}<W<W_{2}$ we get
\begin{eqnarray}
\begin{split}
  &W=W_{1}+(W_{2}-W_{1})\textrm{sn}^{2}(\tau,k), \quad &&\tau=c_{1}\rho^{\frac{2}{3}}+c_{2}, \quad k^{2}=\frac{W_{2}-W_{1}}{W_{3}-W_{1}}\\
  &M= \rho^{-1/3}(\lambda_{0}W)^{1/2},
  \qquad &&c_{1}^{2}=\frac{9a_{0}\lambda_{0}}{8}(W_{3}-W_{1}) \\
  &\Phi=\frac{\delta}{\sqrt{W_{3}-W_{1}}}\int\frac{d\tau}{W}+\Phi_{0},
   \qquad &&\psi=M\exp[i(\pm\frac{1}{3}\theta+\Phi)]\\
\end{split}
\end{eqnarray}
with  $a_{0},\lambda_{0}<0$ for the case (a) and
$a_{0},\lambda_{0}>0$  for the case (b).

(iv) Last we investigate the case where $P(W)$ has two complex
roots and a real root. Let $W_{2}=p+iq$, $W_{3}=p-iq$, then $p,q$
and $W_{1}$ are found from the system of equations
$$2p+W_{1}=0, \qquad \frac{\delta^{2}}{W_{1}}+2pW_{1}=C, \qquad
p^{2}+q^{2}=\frac{\delta^{2}}{W_{1}}.$$ Since $W_{1}>0$,
integration on $W>W_{1}$ provides the following solution for
$a_{0},\lambda_{0}>0$
\begin{eqnarray}
\begin{split}
  &W=\frac{A+W_{1}+(W_{1}-A)\textrm{cn}(\tau,k)}{1+\textrm{cn}(\tau,k)},  \qquad &&\tau=c_{1}\rho^{\frac{2}{3}}+c_{2}, \\
  &M= \rho^{-1/3}(\lambda_{0}W)^{1/2}, \qquad  &&k^{2}=\frac{A+p-W_{1}}{2A},\\
  &\Phi=\frac{\delta}{2\sqrt{A}}\int\frac{d\tau}{W}+\Phi_{0}, \qquad &&c_{1}^{2}=\frac{3Aa_{0}\lambda_{0}}{2}, \\
  &\psi=M\exp[i(\pm\frac{1}{3}\theta+\Phi)], \qquad &&A^{2}=(p-W_{1})^{2}+q^{2}. \\
\end{split}
\end{eqnarray}

\subsubsection{Equation \eqref{22b}, with $C_{0}=0$}
From \eqref{2,33} We have
\begin{equation}\label{2,18}
\dot{W}^2=4W(C+W^{2}), \quad C=\text{const.}
\end{equation}
Depending on the integration constant C, we obtain the following
solutions for \eqref{cse}:

(i) \quad $C=0$. In this case the solution is written as
\begin{equation}
\begin{split}
  \psi=\frac{2}{3}(\frac{2}{a_{0}})^{1/2}(\rho^{2/3}-\rho_{0})^{-1}
  \exp[i(\pm\frac{1}{3}\theta+\Phi_{0})],\quad \Phi_0=\text{Const.}.
\end{split}
\end{equation}

(ii)\quad $C=-p^{2}<0$. We write Eq. \eqref{2,18} as
$$\dot{W}^{2}=4W(W-p)(W+p).$$  Integration for $W>p$ and $-p<W<0$ gives
respectively
\begin{eqnarray}
\begin{split}
    &\psi=\frac{2 c_{1}}{3}(\frac{1}{a_{0}})^{1/2}\rho^{-1/3}\; [\textrm{cn}(c_{1}\rho^{2/3}+c_{2},k)]^{-1}\exp[i(\pm\frac{1}{3}\theta+\Phi_{0})],
    \qquad &&a_{0}, \lambda_{0}>0 \; ;  \\
    &\psi=\frac{2 c_{1}}{3}(\frac{-1}{a_{0}})^{1/2}\rho^{-1/3}\;\textrm{cn}\left(c_{1}\rho^{2/3}+c_{2},k\right)\exp[i(\pm\frac{1}{3}\theta+\Phi_{0})],
    \qquad &&a_{0}, \lambda_{0}<0
\end{split}
\end{eqnarray}
with $c_{1}^{2}=\frac{9a_{0}\lambda_{0}(-C)^{1/2}}{4}$,
$k^{2}=1/2$.

(iii) \quad $C=p^{2}>0$.  The solution  for $a_{0},\lambda_{0}>0$
is given by integrating \eqref{2,18}
\begin{equation}
     \psi=\frac{2 c_{1}}{3}(\frac{1}{a_{0}})^{1/2}\rho^{-1/3}\,\textrm{tn}(c_{1}\rho^{2/3}+c_{2},k)\,
     \textrm{dn}(c_{1}\rho^{2/3}+c_{2},k)\exp[i(\pm\frac{1}{3}\theta+\Phi_{0})]\\
\end{equation}
with $c_{1}^{2}=\frac{9 a_{0}\lambda_{0}C^{1/2}}{8}, \quad
k^{2}=1/2$.

\subsection{Solutions using the subalgebra $L_{2,3}$}
(i) Eq. \eqref{23b} with $C_0\neq0$. If we set
$\lambda=\lambda_0\rho^{-2/3}$ and $\eta=\eta_0\rho^{2/3}$ Eq.
\eqref{23b} becomes

$$\ddot{W}=\frac{1}{2W}\dot{W}^2+\frac{9a_0\lambda_0}{2\eta_0^2}W^2
+\left[\frac{9a^2-1}{2}\eta^{-2}+\frac{9\varepsilon}{2\eta_0^{3}}\eta\right]W+
\frac{9}{2}\left(\frac{C_0}{\lambda_0\eta_0}\right)^2W^{-1}.$$

If we choose $\eta_0=-(\frac{9}{2})^{1/3}\varepsilon$ and
$a^2=1/9$, this equation transforms to an equation quite similar
to PXXXIV
$$\ddot{W}=\frac{1}{2W}\dot{W}^2+4\gamma W^2-\eta W+2\delta^{2}W^{-1}$$
where $\gamma=3^{2/3}2^{-7/3}a_0\lambda_0$,
$\delta^2=3^{2/3}2^{-4/3}(C_0/\lambda_0)^2$. We put \cite{Ince56}
$$2\gamma W=\dot V+V^2+\frac{\eta}{2}$$ and see
that $V$ solves
$$\ddot V=2V^3+\eta V+k, \qquad k=-\frac{1}{2}\pm4a_0\delta i,$$
which is the second Painlev\'e transcendent. Hence we get
$V=P_{II}\left[-(\frac{9}{2})^{1/3}\varepsilon \rho^{2/3}\right]$.
Since $W$ is complex-valued, $\lambda_0$ is a complex constant
which must be chosen appropriately to make $\lambda_0W$ real.

\noindent(ii) Eq. \eqref{23b} with $C_0=0$. We have
$$\ddot{W}=\frac{1}{2W}\dot{W}^2+\frac{9a_0\lambda_0}{2\eta_0^2}W^2
+\left[\frac{9a^2-1}{2}\eta^{-2}+\frac{9\varepsilon}{2\eta_0^{3}}\eta\right]W.$$
With the choice of $a^2=\frac{1}{9}$,
$\eta_0=(\frac{3}{2})^{2/3}\varepsilon$ and
$\lambda_0=(\frac{2}{9})^{5/3}\frac{1}{a_0}$, this equation
transforms to PXX
$$\ddot{W}=\frac{1}{2W}\dot{W}^2+4W^2+2\eta W.$$ Setting $U^2=W$
we again reach to the Painlev\'e transcendent $P_{II}$:
$$\ddot U=2U^3+\eta U. $$
Hence we write the solution as
\begin{equation}
\psi=\lambda_0^{1/2}\rm
P_{II}(\eta_0\rho^{2/3})\exp[i(\pm\frac{1}{3}\theta+\varepsilon
\ln t +\Phi_0)].
\end{equation}

\subsection{Solutions using the subalgebra $L_{2,4}$}
Eq. \eqref{single2} for \eqref{24b} is
\begin{equation}\label{241}
\begin{split}
    \ddot{W}&=\frac{1}{2W}\dot{W}^2-\frac{1}{\dot\eta}(\frac{1}{\xi}+
    \frac{\ddot{\eta}}{\dot{\eta}}
    +\frac{\dot{\lambda}}{\lambda})\dot{W}
    +\frac{a_0\lambda}{2\xi\dot{\eta}^2}W^2+\frac{C_0^2}{32(\xi\lambda{\dot{\eta)}}^2}W^{-1}\\
    &+\frac{1}{\dot{\eta}^2}(\frac{b}{4\xi}+\frac{a^2}{2\xi^2}-\frac{1}{32}-
    \frac{\dot{\lambda}}{\lambda \xi}+
    \frac{\dot{\lambda}^2}{2\lambda^2}-\frac{\ddot{\lambda}}{\lambda})W.
\end{split}
\end{equation}

As  was done for the subalgebra $L_{2,1}$, this equation may be
comparable to PXXXIII or PXXXIV of Ref. \cite{Ince56}. If we
compare the coefficients of $W^2$ and $\dot{W}$  in \eqref{241}
with \eqref{P33} and \eqref{P34} we see that $\lambda$ and $\eta$
must be chosen as
$$\lambda=\lambda_0\xi^{-1/3}, \quad \eta=\eta_0\xi^{1/3}.$$
For this choice of $\lambda$ and $\eta$, a comparison of the
coefficients with those of the possible canonical equations shows
us that it can not have the Painlev\'e property.

If we consider Eq. \eqref{24b} with $C_0=0$, the same arguments
above follows.  We stress that Eq. \eqref{24b} admits no point
symmetries and no constant solutions.

\subsection{Solutions using the subalgebra $L_{2,5}$}
\subsubsection{Equation \eqref{25b}, with $C_{0}\neq0$} By a suitable
choice of $\lambda$ and $\eta$ this equation can be transformed to
a standard form  which has no first derivative and a constant
coefficient of the quadratic term. The transformation and the
corresponding transformed equation are given by
$$\lambda=\lambda_0,\quad
\eta=\eta_0y,$$
$$\ddot{W}=\frac{1}{2W}\dot{W}^2+(\frac{2a_{0}\lambda_0}{\eta_0^2})W^2
+ 2(\frac{C_0}{\lambda_0\eta_0})^{2}W^{-1}.$$ In this case we set
$\eta_0^{2}=a_{0}\lambda_0/2$ and integrate once to obtain the
differential equation for  Weierstrass $\wp(\xi)$ function or
Jacobi elliptic functions \cite{Byrd71}
\begin{equation}\label{5,33}
\dot{W}^2=4(W^3+CW-\delta^{2}), \quad C=\text{const.}, \quad
\delta^2=\frac{2C_0^2}{a_0\lambda_0^3}.
\end{equation}
By virtue of \eqref{posH} we require $\lambda W$ to be positive.
If we factor the cubic polynomial on the right side
$$P(W)=4(W^3+CW-\delta^{2})=4(W-W_1)(W-W_2)(W-W_3),$$
where $W_1, W_2, W_3$ are roots of the cubic polynomial $P(W)$, we
see that there are four possibilities for the solutions of
\eqref{5,33}:

(i)\quad $P(W)$ has only one root. This is not possible since
$P(W)$ does not include the quadratic term $W^{2}$.

(ii) $P(W)$ has a double root, say $W_{1}=W_{2}\neq W_{3}$ and
write $P(W)$ as
$$P(W)=4(W^3+CW-\delta^{2})=4(W-W_1)^{2}(W-W_3).$$
This case is possible for
$C=-3.2^{\frac{-2}{3}}\delta^{\frac{4}{3}}$ and $W_{1,2,3}$ are
given as $W_{1,2}=-2^{-\frac{1}{3}}\delta^{\frac{2}{3}}$,
$W_{3}=(2\delta)^{\frac{2}{3}}$. Integrating for  $W>W_{3}$ we
arrive at the solution given by

\begin{equation}
\begin{split}
 &M= c_{1}(\frac{2}{a_{0}})^{\frac{1}{2}}(-\frac{1}{3}+\sec^{2}\tau)^{\frac{1}{2}},
 \qquad \tau=c_{1}y+c_{2}, \qquad
 c_{1}^{2}=3.2^{-4/3}a_{0}\lambda_{0}\delta^{2/3},\\
 &\Phi=\tan^{-1}(\sqrt{\frac{3}{2}}\tan\tau)-\sqrt{\frac{2}{3}}\tau+\Phi_{0},
 \qquad \psi=M\exp(i\Phi).\\
\end{split}
\end{equation}

(iii)  $P(W)$  has three distinct roots. In this case, roots
$W_{1,2,3}$ are determined by the system
$$W_1+W_2+W_3=0,\quad W_1W_2+W_1W_3+W_2W_3=C,\quad
W_1W_2W_3=\delta^{2}.$$  We see that two of the roots have the
same sign, while the third is always positive. There exist two
orderings of the roots yielding different solutions.
$$(a)\quad  W_{1}<W_{2}<0<W_{3} \qquad (b)\quad 0<W_{1}<W_{2}<W_{3},
\quad W_{2}^{2}<\frac{\delta^{2}}{W_{1}}.$$
 Integration for $W>W_{3}$ gives both for (a) and (b)
\begin{eqnarray}
\begin{split}
  &W=W_{2}+(W_{3}-W_{2})[\textrm{cn}(\tau,k)]^{-2}, \quad &&\tau=c_{1}y+c_{2},
  \quad k^{2}=\frac{W_{2}-W_{1}}{W_{3}-W_{1}},\\
  &M= (\lambda_{0}W)^{1/2},
  \qquad &&c_{1}^{2}=\frac{a_{0}\lambda_{0}}{2}(W_{3}-W_{1}), \\
  &\Phi=\frac{\delta}{\sqrt{W_{3}-W_{1}}}\int\frac{d\tau}{W}+\Phi_{0},
   \qquad &&\psi=M\exp(i\Phi)\\
\end{split}
\end{eqnarray}
with $a_{0},\lambda_{0}>0$.

Integrating on $W_{1}<W<W_{2}$ we get
\begin{eqnarray}
\begin{split}
  &W=W_{1}+(W_{2}-W_{1})\textrm{sn}^{2}(\tau,k), \qquad
  &&\tau=c_{1}y+c_{2},\quad
  k^{2}=\frac{W_{2}-W_{1}}{W_{3}-W_{1}}, \\
  &M=(\lambda_{0}W)^{1/2},\quad
  &&c_{1}^{2}=\frac{a_{0}\lambda_{0}}{2}(W_{3}-W_{1}),  \\
  &\Phi=\frac{\delta}{\sqrt{W_{3}-W_{1}}}\int\frac{d\tau}{W}+\Phi_{0},\quad
  &&\psi=M\exp(i\Phi)
\end{split}
\end{eqnarray}
with  $a_{0},\lambda_{0}<0$ for the case (a) and
$a_{0},\lambda_{0}>0$  for the case (b).

(iv) Finally, we investigate the case where $P(W)$ has two complex
roots and a real root. Let $W_{2}=p+iq$, $W_{3}=p-iq$, then $p,q$
and $W_{1}$ are found from the system of equations $$2p+W_{1}=0,
\qquad \frac{\delta^{2}}{W_{1}}+2pW_{1}=C, \qquad
p^{2}+q^{2}=\frac{\delta^{2}}{W_{1}}.$$ Since $W_{1}>0$,
integration on $W>W_{1}$ gives the following solution for
$a_{0},\lambda_{0}>0$
\begin{eqnarray}
\begin{split}
  &W=\frac{A+W_{1}+(W_{1}-A)\textrm{cn} (\tau,k)}{1+\textrm{cn} (\tau,k)},  \qquad &&\tau=c_{1}y+c_{2}, \\
  &M= (\lambda_{0}W)^{1/2}, \qquad  &&k^{2}=\frac{A+p-W_{1}}{2A}\\
  &\Phi=\frac{\delta}{2\sqrt{A}}\int\frac{d\tau}{W}+\Phi_{0}, \qquad &&c_{1}^{2}=2Aa_{0}\lambda_{0} \\
  &\psi=M\exp(i\Phi), \qquad &&A^{2}=(p-W_{1})^{2}+q^{2}. \\
\end{split}
\end{eqnarray}

\subsubsection{Equation \eqref{25b}, with $C_{0}=0$}
From \eqref{5,33} we have
\begin{equation}\label{5,18}
\dot{W}^2=4W(C+W^{2}), \quad C=\text{const.}
\end{equation}
Depending on the integration constant C, we obtain the following
elliptic solutions for \eqref{cse}:

(i) \quad $C=0$. In this case the solution is written for
$a_{0}>0$ :
\begin{equation}\label{}
\begin{split}
  \psi=(\frac{2}{a_{0}})^{1/2}(y-y_{0})^{-1}\exp(i\Phi_{0}).
\end{split}
\end{equation}

(ii)\quad $C=-p^{2}<0$. We write Eq. \eqref{5,18} as
$$\dot{W}^{2}=4W(W-p)(W+p).$$  Integration for $W>p$ and $-p<W<0$ gives
respectively
\begin{equation}
\begin{split}
    &\psi=c_{1}(\frac{1}{a_{0}})^{1/2}\;[\textrm{cn}(c_{1}y+c_{2},k)]^{-1}\exp(i\Phi_{0}),
    \qquad a_{0}, \lambda_{0}>0 \; ; \\
    &\psi=c_{1}(\frac{-1}{a_{0}})^{1/2}\;\textrm{cn}\left(c_{1}y+c_{2},k\right)\exp(i\Phi_{0}),
    \qquad\quad\, a_{0}, \lambda_{0}<0
\end{split}
\end{equation}
with $c_{1}^{2}=a_{0}\lambda_{0}(-C)^{1/2}$, $ k^{2}=1/2$.

(iii) \quad $C=p^{2}>0$.  The solution  for $a_{0},\lambda_{0}>0$
is given by integrating \eqref{5,18}\; :
\begin{equation}\label{}
     \psi=c_{1}(\frac{2}{a_{0}})^{1/2}\,\textrm{tn}(c_{1}y+c_{2},k)\,
     \textrm{dn}(c_{1}y+c_{2},k)\exp(i\Phi_{0})
\end{equation}
with $c_{1}^{2}= \frac{a_{0}\lambda_{0}C^{1/2}}{2},$  $k^{2}=1/2$.

\subsection{Solutions using the subalgebra $L_{2,6}$}
\subsubsection{Equation \eqref{26b}, with $C_{0}\neq0$} By a suitable
choice of $\lambda$ and $\eta$ this equation can be transformed to
a standard form  which has no first derivative and a constant
coefficient of the quadratic term. The transformation and the
corresponding transformed equation are given by
$$\lambda=\lambda_0,\quad
\eta=\eta_0y, $$
$$\ddot{W}=\frac{1}{2W}\dot{W}^2+(\frac{2a_0\lambda_0}{\eta_0^2})W^2+\frac{2\varepsilon}{\eta_{0}^{2}}W
+ 2(\frac{C_0}{a_0\lambda_0})^{2}W^{-1}.$$ In this case we set
$\eta_0^{2}=a_{0}\lambda_0/2$ and integrate once to obtain
\begin{equation}\label{6,33}
\dot{W}^2=4(W^3+\alpha W^{2}+CW-\delta^{2}), \quad
C=\text{const.},
\end{equation}
where $\delta^2={2C_0^2}/(a_0\lambda_0^3)$ and $\alpha
=2\varepsilon/(a_{0}\lambda_{0})$. Solution of this equation can
be expressed in terms of Weierstrass $\wp(\xi)$ function or Jacobi
elliptic functions. By virtue of \eqref{posH} we require $\lambda
W$ to be positive. Let us write the right side as
$$P(W)=4(W^3+\alpha W^{2}+CW-\delta^{2})=4(W-W_1)(W-W_2)(W-W_3),$$
where $W_1, W_2, W_3$ are roots of the cubic polynomial $P(W)$. We
then have to distinguish between four different cases:

(i)\quad $P(W)$ has only one root. This is the case when
$$W_{1}=\frac{-\alpha}{3}=(\frac{C}{3})^{1/2}=\delta^{2/3}.$$
The solution is
\begin{equation}
\begin{split}
 &M=(\frac{2\varepsilon}{3a_{0}}+\frac{\lambda_{0}}{\tau})^{1/2},
 \qquad \tau^2=c_{1}y+c_{2}, \qquad
 c_{1}^{2}=8a_{0}\lambda_{0},\\
 &\Phi=\frac{3\delta}{4\alpha}\tau^{2}-\frac{9\delta}{2\alpha^{2}}\tau+\frac{27\delta}{2\alpha^{3}}\ln(3+\alpha\tau)+\Phi_{0},
 \qquad \psi=M\exp[i(\varepsilon t+\Phi)].\\
\end{split}
\end{equation}

(ii) $P(W)$ has a double root, say $W_{1}=W_{2}\neq W_{3}$ and
write $P(W)$ as
$$P(W)=4(W^3+\alpha W^{2}+CW-\delta^{2})=4(W-W_1)^{2}(W-W_3).$$
We find $W_{1}$ and $W_{3}$ from the system
$$-\alpha=W_{3}+2W_{1} \quad C=2W_{1}W_{3}+W_{1}^{2} \quad
\delta^{2}=W_{3}W_{1}^{2}.$$ The roots are ordered as follows:
$$(a)\quad W_{1}>\delta^{2/3} \quad \Rightarrow \quad
0<W_{3}<W_{1},\quad (b)\quad W_{1}<\delta^{2/3}\quad\Rightarrow
\quad W_{1}<W_{3}.$$ For the case (a) we integrate on $W>W_{1}$
and $W_{3}<W<W_{1}$, then find
$$W=W_{1}+(W_{1}-W_{3})\textrm{cosech}^{2}\tau \qquad
\text{and} \qquad W=W_{1}+(W_{3}-W_{1})\textrm{sech}^{2}\tau $$
respectively. The corresponding solutions are
\begin{equation}
\begin{split}
  &M=(\lambda_{0}W)^{1/2}, \qquad \tau=c_{1}y+c_{2},
  \quad c_{1}^{2}=\frac{a_{0}\lambda_{0}}{2}(W_{1}-W_{3}),\\
  &\Phi=\frac{\delta}{\sqrt{W_{1}-W_{3}}}\int\frac{d\tau}{W}+\Phi_{0}
  \qquad \psi=M\exp[i(\varepsilon t+ \Phi)].
\end{split}
\end{equation}
For the case (b), we integrate on $W>W_{3}$ and obtain the
solution
\begin{eqnarray}
\begin{split}
  &W=W_{1}+(W_{3}-W_{1})\sec^{2}\tau,  \qquad &&\tau=c_{1}y+c_{2}, \\
  &M=(\lambda_{0}W)^{1/2},     \qquad &&c_{1}^{2}=\frac{a_{0}\lambda_{0}}{2}(W_{3}-W_{1}),\\
  &\Phi=\frac{\delta}{\sqrt{W_{3}-W_{1}}}\int\frac{d\tau}{W}+\Phi_{0} \qquad &&\psi=M\exp[i(\varepsilon t+
  \Phi)].
\end{split}
\end{eqnarray}

(iii)  $P(W)$  has three distinct roots. In this case, roots
$W_{1,2,3}$ are determined by the system
$$W_1+W_2+W_3=-\alpha,\quad W_1W_2+W_1W_3+W_2W_3=C,\quad
W_1W_2W_3=\delta^{2}.$$ We see that two of the roots have the same
sign, while the third is always positive. There are two cases to
consider:
$$(a)\quad W_{1}<W_{2}<0<W_{3} \qquad (b)\quad 0<W_{1}<W_{2}<W_{3}, \quad
W_{2}^{2}<\frac{\delta^{2}}{W_{1}}.$$ Integration for $W>W_{3}$
gives both for (a) and (b)
\begin{eqnarray}
\begin{split}
  &W=W_{2}+(W_{3}-W_{2})[\textrm{cn}(\tau,k)]^{-2}, \quad &&\tau=c_{1}y+c_{2},
  \quad k^{2}=\frac{W_{2}-W_{1}}{W_{3}-W_{1}},\\
  &M= (\lambda_{0}W)^{1/2},
  \qquad &&c_{1}^{2}=\frac{a_{0}\lambda_{0}}{2}(W_{3}-W_{1}), \\
  &\Phi=\frac{\delta}{\sqrt{W_{3}-W_{1}}}\int\frac{d\tau}{W}+\Phi_{0},
   \qquad &&\psi=M\exp[i(\varepsilon t+\Phi)]\\
\end{split}
\end{eqnarray}
with $a_{0},\lambda_{0}>0$. Integrating on $W_{1}<W<W_{2}$ we get
\begin{eqnarray}
\begin{split}
  &W=W_{1}+(W_{2}-W_{1})\textrm{sn}^{2}(\tau,k), \qquad &&\tau=c_{1}y+c_{2},\quad k^{2}=\frac{W_{2}-W_{1}}{W_{3}-W_{1}}, \nonumber\\
  &M= (\lambda_{0}W)^{1/2}, \qquad  &&c_{1}^{2}=\frac{a_{0}\lambda_{0}}{2}(W_{3}-W_{1}), \nonumber \\
  &\Phi=\frac{\delta}{\sqrt{W_{3}-W_{1}}}\int\frac{d\tau}{W}+\Phi_{0}, \qquad &&\psi=M\exp[i(\varepsilon t+\Phi)]
\end{split}
\end{eqnarray}
with  $a_{0},\lambda_{0}<0$ for the case (a) and
$a_{0},\lambda_{0}>0$  for the case (b).

(iv) Last we investigate the case where $P(W)$ has two complex
roots and a real root. Let $W_{2}=p+iq$, $W_{3}=p-iq$, then $p,q$
and $W_{1}$ are found from the system of equations
$$2p+W_{1}=\alpha, \qquad \frac{\delta^{2}}{W_{1}}+2pW_{1}=C,
\qquad p^{2}+q^{2}=\frac{\delta^{2}}{W_{1}}.$$ Since $W_{1}>0$,
integration on $W>W_{1}$ gives  the following solution for
$a_{0},\lambda_{0}>0$.
\begin{eqnarray}
\begin{split}
  &W=\frac{A+W_{1}+(W_{1}-A)\textrm{cn} (\tau,k)}{1+\textrm{cn} (\tau,k)},  \qquad &&\tau=c_{1}y+c_{2}, \\
  &M= (\lambda_{0}W)^{1/2}, \qquad  &&k^{2}=\frac{A+p-W_{1}}{2A},\\
  &\Phi=\frac{\delta}{2\sqrt{A}}\int\frac{d\tau}{W}+\Phi_{0}, \qquad &&c_{1}^{2}=2Aa_{0}\lambda_{0}, \\
  &\psi=M\exp[i(\varepsilon t+\Phi)], \qquad &&A^{2}=(p-W_{1})^{2}+q^{2}. \\
\end{split}
\end{eqnarray}

\subsubsection{Equation \eqref{26b} with $C_{0}=0$}
In this case, Eq. for $W$ reduces to
\begin{equation}\label{6,19}
\ddot{W}=\frac{1}{2W}\dot{W}^2+(\frac{2a_{0}\lambda_0}{\eta_0^2})W^2+\frac{2\varepsilon}{\eta_{0}^{2}}W.
\end{equation}
We are going to investigate the solutions of Eq. \eqref{6,19} for
$\varepsilon=1$ and $\varepsilon=-1$ separately.  By the choice
$\eta_0^{2}=\frac{a_{0}\lambda_0}{2}=1$ this equation transforms
to
$$\ddot{W}=\frac{1}{2W}\dot{W}^2+4W^2+2\varepsilon W$$
which is PXIX of \cite{Ince56} for $\varepsilon=1$. The first
integral is again the elliptic function equation
\begin{equation}\label{6,19c}
\dot{W}^2=4W(C+\varepsilon W+W^{2}), \quad C=\text{const.}
\end{equation}
If we label the right-hand side of  Eq. \eqref{6,19c} as $P(W)$ we
find the roots of the cubic polynomial $P(W)$ to be
$$W_{1}=\frac{-1-\sqrt{1-4C}}{2}, \quad W_{2}=0, \quad
W_{3}=\frac{-1+\sqrt{1-4C}}{2} \quad \text{for}\quad
\varepsilon=1,$$ and
$$W_{4}=\frac{\,1-\sqrt{1-4C}}{2}, \quad W_{5}=0, \quad
W_{6}=\frac{\,1+\sqrt{1-4C}}{2} \quad \text{for}\quad
\varepsilon=-1.$$ We see that ordering of the roots depends on  C:
\begin{equation}
\begin{split}
C<0 \quad &\Rightarrow  \quad W_{1,4}<0<W_{3,6}\quad \text{for}      \quad
\varepsilon=1, \\
0<C<\frac{1}{4} \quad &\Rightarrow  \quad 0<W_{1,4}<W_{3,6}\quad \text{for}  \quad
\varepsilon=-1.
\end{split}
\end{equation}
Depending on the integration constant C, we obtain the following
solutions for \eqref{cse}:

(i) \quad $C=0$. In this case the solution  is given below with
$\kappa=\pm1$ for suitable sign of $a_0$.
\begin{equation}
\begin{split}
  0<W,\quad &\psi=(\frac{2}{a_{0}})^{1/2}\textrm{cosech} (\kappa y+c_{1})\exp[i(t+\Phi_{0})] \; ;\\
  -1<W<0,\quad &\psi=(\frac{-2}{a_{0}})^{1/2}\textrm{sech}(\kappa y+c_{1})\exp[i(t+\Phi_{0})] \ ;\\
  1<W,\quad  &\psi=(\frac{2}{a_{0}})^{1/2}\sec(\kappa y+c_{1})\exp[i(-t+\Phi_{0})] \; .
\end{split}
\end{equation}

(ii)\quad $C=1/4$.
\begin{equation}
\begin{split}
0<W,\quad &\psi=(\frac{1}{a_{0}})^{1/2}\tan (\frac{\kappa}{\sqrt{2}}y+c_{1})\exp[i(t+\Phi_{0})] \; ;  \\
0<W<\frac{1}{2},\quad &\psi=(\frac{1}{a_{0}})^{1/2}\tanh(\frac{\kappa}{\sqrt{2}}y+c_{1})\exp[i(-t+\Phi_{0})] \; ;\\
\frac{1}{2}<W,\quad
&\psi=(\frac{1}{a_{0}})^{1/2}\textrm{coth}(\frac{\kappa}{\sqrt{2}}y+c_{1})\exp[i(-t+\Phi_{0})]
\end{split}
\end{equation}
with $a_0>0,$  $\kappa=\pm1$.

(iii) \quad $C<0$. For both $\varepsilon=\pm1$, the roots of
$P(W)$ have opposite signs. An integration with respect to
$W_{3,6}<W$ and $W_{1,4}<W<W_{2,5}{}$ results in following
solutions:
\begin{equation}
\begin{split}
&\psi=(\frac{\sqrt{1-4C}-\varepsilon}{a_{0}})^{1/2}\,[\textrm{cn}(c_{1}y+c_{2},k)]^{-1}\exp[i(\varepsilon t+\Phi_{0})],  \quad a_{0}>0\; ;\\
&\psi=(\frac{\sqrt{1-4C}+\varepsilon}{-a_{0}})^{1/2}\,{\rm
cn}(c_{1}y+c_{2},k)\exp[i(\varepsilon t+\Phi_{0})],  \qquad
a_{0}<0
\end{split}
\end{equation}
with $c_{1}^{2}=\sqrt{1-4C},$
$k^{2}=\frac{\varepsilon+\sqrt{1-4C}}{2\sqrt{1-4C}}$.

(iv)\quad $0<C<\frac{1}{4}$. We list the solutions following the
interval that the integration has been taken on:
\begin{equation}
\begin{split}
&W_{2}<W, \\
& \psi=c_{1}\Bigl(\frac{2}{a_{0}}\Bigr)^{1/2}\Bigl(\frac{1-\sqrt{1-4C}}{1+\sqrt{1-4C}}\Bigr)^{1/2}\textrm{tn}(c_{1}y+c_{2},k)\exp[i(t+\Phi_{0})], \qquad a_0>0\; ;  \\
&W_{1}<W<W_{3}, \\
&\psi=c_{1}\Bigl(\frac{-2}{a_{0}}\Bigr)^{1/2}[\frac{1-\sqrt{1-4C}}{1+\sqrt{1-4C}}+\cn^{2}(c_{1}y+c_{2},k)]^{1/2}\exp[i(t+\Phi_{0})],
\quad a_0<0
\end{split}
\end{equation}
with $c_{1}^{2}=\frac{1+\sqrt{1-4C}}{2},$
$k^{2}=\frac{2\sqrt{1-4C}}{1+\sqrt{1-4C}}$.
\begin{equation}
\begin{split}
&W_{6}<W, \\
&\psi=(\frac{1-\sqrt{1-4C}}{a_{0}})^{1/2}[1+\frac{2\sqrt{1-4C}}{1-\sqrt{1-4C}} (\textrm{cn}(c_{1}y+c_{2},k))^{-2}]^{1/2}\exp[i(-t+\Phi_{0})]\; ; \\
&W_{5}<W<W_{4}, \\
&\psi=(\frac{1-\sqrt{1-4C}}{a_{0}})^{1/2}\textrm{sn}(c_{1}y+c_{2},k)\exp[i(-t+\Phi_{0})]
\end{split}
\end{equation}
with $a_{0}>0, \quad c_{1}^{2}=\frac{1+\sqrt{1-4C}}{2},$
$k^{2}=\frac{1-\sqrt{1-4C}}{1+\sqrt{1-4C}}$.

(v)\quad $\frac{1}{4}<C$. We have two solutions for $a_{0}>0$\;:
\begin{equation}
   \psi=(\frac{2}{a_{0}})^{1/2}C^{1/4}\textrm{tn}(C^{1/4}y+c_{1},k) \,\textrm{dn}(C^{1/4}y+c_{1},k)\exp[i(\varepsilon
   t+\Phi_{0})]
\end{equation}
with $k^{2}=\frac{2C^{1/2}-\varepsilon}{4C^{1/2}}.$

\subsection{Solutions using the subalgebra $L_{2,7}$}

This subalgebra leads to second order non-linear ODEs of which
solutions are written in terms of Painlev\'e transcendents.

(i) Eq. \eqref{27b} with $C_0\neq0$.  If we set $\lambda=\lambda_0,
\eta=\eta_0\xi, $  this equation transforms to
$$\ddot{W}=\frac{1}{2W}\dot{W}^2+(\frac{2a_{0}\lambda_0}{\eta_0^2})W^2+\frac{\eta}{\eta_{0}^{3}}W+2(\frac{C_0}{\lambda_0\eta_0})^2W^{-1}
.$$
 For the choice $\eta_0=-1$ this equation becomes
$$\ddot{W}=\frac{1}{2W}\dot{W}^2+4\gamma W^2-\eta W+2\delta^{2}W^{-1},$$
where $\gamma=a_{0}\lambda_0/2$ and $\delta^2=(C_0/\lambda_0)^2$.
This equation is quite similar to equation PXXXIV of \cite{Ince56}
and is integrated in the same way. We put
$$2\gamma W=\dot V+V^2+\frac{\eta}{2}$$
and see that $V$ is solved by the equation
$$\ddot V=2V^3+\eta V+k, \qquad k=-\frac{1}{2}\pm4a_0\delta i,$$
which is the second Painlev\'e transcendent. Hence we find
$V=P_{II}(-\xi)$. Since $W$ is complex-valued, $\lambda_0$ is a
complex constant which must be chosen appropriately to make
$\lambda_0W$ real.

(ii)  Eq. \eqref{27b} with $C_0=0$. Under the same transformation
$\lambda=\lambda_0$, $ \eta=\eta_0\xi, $  this equation reduces to
$$\ddot{W}=\frac{1}{2W}\dot{W}^2+(\frac{2a_{0}\lambda_0}{\eta_0^2})W^2+\frac{\eta}{\eta_{0}^{3}}W.$$
With choice $\eta_0=2^{-1/3}$ and  $\lambda_0=2^{1/3}/a_0$  it is
nothing but the canonical Eq. PXX of \cite{Ince56}
$$\ddot{W}=\frac{1}{2W}\dot{W}^2+4W^2+2\eta W.$$
If we set $U^2=W$,  the second Painlev\'e transcendent $P_{II}$
makes its appearance again
$$\ddot U=2U^3+\eta U. $$
We can write the solution as
\begin{equation}
\psi=\frac{2^{1/6}}{a_{0}^{1/2}}\mathrm{P_{II}}(2^{-1/3}\xi)\exp[i(\frac{xt}{2}-\frac{t^3}{6}+\Phi_0)].
\end{equation}

\subsection{Solutions using the subalgebra $L_{2,10}$}
Eq. \eqref{single2} for \eqref{30b} is
\begin{equation}\label{2101}
\begin{split}
    \ddot{W}&=\frac{1}{2W}\dot{W}^2-\frac{1}{\dot\eta}(\frac{\ddot{\eta}}{\dot{\eta}}
    +\frac{\dot{\lambda}}{\lambda})\dot{W}
    +\frac{2a_0\lambda}{\dot{\eta}^2}W^2+2(\frac{C_0}{\lambda\dot{\eta}})^2W^{-1}\\
    &+\frac{1}{\dot{\eta}^2}(2\xi^2-2a\varepsilon+\frac{\dot{\lambda}^2}{2\lambda^2}-\frac{\ddot{\lambda}}{\lambda})W.
\end{split}
\end{equation}
Just as  for the subalgebra $L_{2,1}$, this equation seems a
candidate to be transformable to PXXXIII or PXXXIV. By comparison
of the coefficients of $W^2$ and $\dot{W}$  in \eqref{2101} with
\eqref{P33} and \eqref{P34} it turns out  that $\lambda$ and
$\eta$ must be chosen to be

$$\lambda=\lambda_0, \quad \eta=\eta_0\xi.$$
For this choice of $\lambda$ and $\eta$ \eqref{2101} becomes
\begin{equation}\label{2102}
\ddot{W}=\frac{1}{2W}\dot{W}^2+\frac{2a_0\lambda_0}{\eta_0^2}W^2+2(\frac{C_0}{\lambda_0\eta_0})^2W^{-1}
+\left(\frac{2}{\eta_0^4}\eta^{2}-\frac{2a\varepsilon}{\eta_0^2}\right)W.
\end{equation}
Since the coefficient of $W$ includes the term $\eta^2$,  Eq.
\eqref{30b} cannot be transformed to one of the canonical forms,
thus it does not have the Painlev\'e property.

If we consider Eq. \eqref{30b} with $C_0=0$, the same arguments
above apply. We finally state that Eq. \eqref{30b} admits no point
symmetries and no constant solutions.

\subsection{Solutions using the subalgebra $L_{2,12}$}
Singularity analysis and integration of Eq. \eqref{32b} are
performed in Ref.\cite{Gagnon89-3} in which the authors  showed
that Eq. \eqref{32b} for $C_0=0$, namely
$$2\dot{\chi}\dddot{\chi}-\ddot{\chi}^2+(\frac{\xi^2}{4}-2a)\dot{\chi}^2-\frac{1}{4}
\chi^2-4a_0\dot{\chi}^3=0$$ is of Painlev\'e type and allows  a
first integral of the form
\begin{equation} \label{2124}
\ddot{\chi}^2=-\frac{1}{4}(\chi-\dot{\chi}\xi)^2+2a\dot{\chi}^2+2a_0+\dot{\chi}^3+C\dot{\chi}.
\end{equation}
They also showed that the solution of \eqref{2124} is given by
\begin{equation} \label{2125}
\chi(\xi)=-\frac{k}{a_0}\left[\frac{\dot{W}^2}{4W}-\frac{W^3}{4}-zW^2+(1-z^2+\alpha)W+\frac{\beta}{2W}
       +\frac{2}{3}(\alpha+1)z+\frac{az}{3k^2}\right],
\end{equation}
where $W(z)$ satisfies the Painlev\'e transcendent $P_{IV}$
\begin{equation} \label{2126}
\ddot{W}=\frac{1}{2W}\dot{W}^2+\frac{3}{2}W^3+4zW^2+2(z^2-\alpha)W+\frac{\beta}{W},
\end{equation}
where $z=k\xi,$  $k^4=-1/16$ and the constants $\alpha, \; \beta$
are determined by
\begin{eqnarray}
\nonumber &&-2(\alpha+1)^2+3\beta=4(2a^2-3Ca_0),\\
\nonumber &&(\alpha+1)[2(\alpha+1)^2+9\beta]=(a/k^2)(4a^2-9Ca_0).
\end{eqnarray}
To sum up, the solution to \eqref{cse} is given by the formula
\eqref{32} with
$$M=\sqrt{\dot{\chi}}, \qquad \Phi=\int\frac{\xi(1+\dot{\chi})}{4\dot{\chi}}\rm{d}\xi.$$

\noindent{\bf Remark:} We have not been able to successfully treat
the reduced equation corresponding to the subalgebra $L_{2,11}$.

\begin{table}[b]\caption{Three-dimensional subalgebras and invariant solutions $(\varepsilon=\pm 1)$}\label{Tab3}
\begin{center}
\begin{tabular}{|c|c|c|}
\hline    No    & Basis & Solution \\
\hline
$L_{3,1}$       & $T+\varepsilon E,P_{1},P_{2}$   & $\psi=(\varepsilon /a_0)^{1/2}\exp[i(\varepsilon t+\Phi_0)]$  \\
 $L_{3,2}$       & $J+aE, D+bE,T$                  & $\psi=[a(x^{2}+y^{2})]^{-1/2}\exp[i(a_0\tan^{-1}(y/x)+\Phi_0)] $ \\
 $L_{3,3}$       & $D+aE, T ,P_{1}$                & $\psi=(\sqrt{2}/y)\exp(i\Phi_0) $ \\
$L_{3,4}$       & $J+\varepsilon T+aE, P_1, P_2$  & $\psi=(a/\varepsilon)^{1/2}\exp[i(at/\varepsilon+\Phi_0)]$  \\
\hline
\end{tabular}
\end{center}
\end{table}

\subsection{Invariant solutions from three-dimensional subalgebras}
We conclude by looking for further solutions on the basis of three
dimensional subalgebras. Invariance under those subalgebras gives
rise to algebraic equations only rather than ODEs. We simply list
nontrivial solutions in table \ref{Tab3}, ruling out those leading
to trivial ones.


\section{Concluding remarks}

In this paper we considered a cubic Schr\"{o}dinger equation in
2+1-dimension and gave a comprehensive investigation of
group-invariant solutions by performing  reductions to  ODEs using
appropriate subgroups of the symmetry group and by solving them
when possible. Our approach is a powerful combination of the group
theory and singularity analysis. In this respect, it is very
similar to the one used in a series of papers for an investigation
of group-invariant solutions of a 3+1-dimensional generalized
nonlinear Schr\"{o}dinger equation (which, as special cases, contains
dissipative, cubic and quintic Schr\"{o}dinger equations) in
\cite{Gagnon88, Gagnon89-2, Gagnon89-3}. The solutions obtained
represent a conjugacy class of solutions. The entire class can be
produced by applying a general symmetry group transformation
\eqref{global}-\eqref{conf} to the representative solution. The
solutions include Jacobi elliptic functions, their degenerate
forms as elementary ones and Painlev\'e irreducibles. What emerges
from a comparison of results in 3+1- and 2+1-dimensions is that
there are reductions to  ODEs with similar structure in both
dimensions (for instance, Painlev\'e transcendents are ubiquitous)
while the corresponding invariant solutions have truly different
expressions.

We mention that some similarity solutions of \eqref{cse} have
already been discussed in \cite{Tajiri83} in which the author
merely chooses an arbitrary combination of the generators for
performing a symmetry reduction. Reduction to an ODE is achieved
in two successive steps, first to a 1+1-dimensional PDE and then
to an ODE by imposing an inherited symmetry of the reduced PDE. On
the contrary, in the present article  we classify solutions into a
list of conjugacy classes by using a rigorous subgroup
classification so that every solution is conjugate to precisely
one in the list.

\subsection*{Acknowledgements} The authors thank the referees for
their constructive and useful comments which substantially
improved the presentation of the paper.


\end{document}